\documentclass[9pt,conference]{IEEEtran}
\usepackage{dcase2026}

\usepackage{bm} 
\usepackage{comment}
\usepackage{multirow}

\usepackage{dcase2026,amsmath,graphicx,url,times,booktabs, tabularx}

\title{RealDESED: A Real-World Domestic Sound Event Detection Benchmark}

\name{Florian Schmid$^{1*}$\thanks{$^*$These authors contributed equally to this work. \\
The LIT AI Lab is supported by the Federal State of Upper Austria.
Gerhard Widmer's work is supported by the European Research Council
under the European Union's Horizon 2020 research and innovation
programme (Grant No.\ 101019375, Whither Music?).},
      Paul Primus$^{1*}$,
      Alexander Fichtinger$^{1}$,
      Tara Jadidi$^{1}$,
      Tobias Morocutti$^{1}$,
      Gerhard Widmer$^{1,2}$}
\address{$^1$Institute of Computational Perception, Johannes Kepler University Linz, Austria \\
$^2$LIT Artificial Intelligence Lab, Linz, Austria\\
first.last@jku.at\\
}

\begin{document}

\maketitle

\begin{abstract}
This paper presents RealDESED, a real-world domestic sound event detection (SED) benchmark comprising 5,710 audio recordings collected by 652 participants in their homes. Each recording is between 15 and 35 seconds long and contains temporally precise annotations for 15 common domestic sound classes. In contrast to existing SED datasets, which typically rely on simulated soundscapes or broad web-crawled audio, RealDESED consists exclusively of recordings captured in natural domestic environments, reflecting realistic variability in recording devices, device placement, acoustic conditions, background sounds, and naturally occurring event co-occurrences. A distinguishing characteristic of the dataset is its multi-annotator labeling scheme, where each recording is independently annotated by multiple annotators, while the validation and test sets undergo an additional review process to ensure high annotation quality and reliable benchmarking. Furthermore, the dataset provides rich metadata, including recording device, device placement, environment labels, and textual scene descriptions. We establish a strong transformer-based baseline and investigate annotation aggregation strategies, post-processing methods, long-form inference, and the impact of recording metadata on model performance. Our baseline achieves a macro-averaged PSDS1 score of 0.731 on the test set. We believe RealDESED provides a valuable benchmark for developing and evaluating robust SED systems under realistic domestic conditions, helping to bridge the gap between current research benchmarks and real-world deployment.
\end{abstract}

\begin{IEEEkeywords}
sound event detection, domestic environment recordings, real-world dataset, multi-annotator labeling, temporal event localization
\end{IEEEkeywords}

\section{Introduction}
\label{sec:intro}

The goal of Sound Event Detection (SED) is to identify acoustic events and their temporal boundaries within audio recordings. Reliable SED systems enable applications such as security and surveillance~\cite{radhakrishnan2005audio}, smart homes and assistive technologies~\cite{debes2016monitoring,homesound}, and health monitoring~\cite{zigel2009method}. Advances in deep learning have substantially improved SED performance over the past decade, making SED an important component of intelligent systems that interpret acoustic environments.

\textbf{Existing SED Benchmarks:} Progress in SED has been driven by publicly available datasets and benchmarks, enabling reproducible development and evaluation. The largest strongly annotated benchmark is \textit{AudioSet Strong}~\cite{audioset,audioset_strong}, comprising about 100k web-crawled recordings spanning 447 sound classes. Other SED datasets are typically an order of magnitude smaller and focus on specific domains. Among these, \textit{DESED}~\cite{desed_turpault,desed_serizel} is the dominant benchmark for domestic SED, combining synthetic training soundscapes with strongly annotated web-crawled recordings for evaluation. Similarly, \textit{UrbanSED}~\cite{scaper} consists of synthetic urban soundscapes with automatically derived strong labels, while \textit{MAESTRO Real}~\cite{maestro_real} provides long-form recordings with strong labels inferred from human-annotated weak labels, limiting temporal annotation granularity.

\textbf{Current Limitations:} Existing SED datasets have several limitations. First, obtaining temporally precise annotations is expensive, resulting in relatively small strongly annotated datasets. For example, while weakly labeled AudioSet contains about 2M recordings, its strongly annotated subset is around 20 times smaller. Second, temporal annotations are inherently subjective, particularly regarding event continuity and onset and offset boundaries, motivating multiple independent annotations for robust label aggregation. A common alternative is synthetic soundscape generation with automatically derived strong labels, as used in DESED and UrbanSED. However, this introduces a domain gap between synthetic and real recordings, often reducing performance on real-world audio. Finally, many datasets rely on web-crawled recordings and fixed 10-second clips, limiting realism and temporal context.

\textbf{Research Directions in SED:} These limitations have strongly influenced SED research. To reduce the reliance on expensive strong annotations, researchers have explored learning from weak labels~\cite{KongMil,desed_turpault,XuAttentionPooling}, semi-supervised learning~\cite{desed_turpault}, heterogeneous supervision~\cite{SchmidDCASE24Task4,NamDCASE24Task4,CornellDCASE24Task4}, and varying pre-training paradigms~\cite{PretrainedSED,matsed,pmam,jigsaw}. Together, these approaches improve data efficiency but remain constrained by the limited availability of realistic strongly annotated benchmark datasets.


\textbf{RealDESED:} In this paper, we present \textit{RealDESED}, a new dataset for domestic sound event detection that addresses the limitations of existing benchmarks. Compared with existing datasets, \textit{RealDESED} offers the following key characteristics:



\begin{itemize}
    \item \textbf{Real-world domestic recordings} collected by 652 participants in their homes, capturing realistic variability in recording devices, environments, and recording conditions, instead of relying on web-crawled recordings or simulated audio mixtures.
    
    \item \textbf{Longer recordings} of 15--35 seconds, providing longer acoustic context and compound action sequences than the fixed 10-second clips commonly used in existing SED benchmarks.
    
    \item \textbf{A benchmark of 5,710 real recordings} with predefined training, validation, and test splits, enabling system development and evaluation on real in-distribution data.
    
    \item \textbf{Multi-annotator temporal annotations}, with reviewed validation and test labels.
    
    \item \textbf{Rich recording metadata}, including recording device, device placement, environment labels, and scene descriptions.
\end{itemize}

We establish a strong transformer-based baseline using ATST-F~\cite{atstf}, pre-trained on AudioSet Strong~\cite{PretrainedSED} and fine-tuned on RealDESED. We further investigate annotation aggregation, long-form inference on recordings exceeding the model's input length, and the impact of recording metadata on system performance, establishing a comprehensive benchmark for future research. To facilitate reproducible research, we publicly release the dataset, baseline implementation, and data collection and annotation guidelines.\footnote{%
  \url{https://github.com/fschmid56/RealDESED}\par
 \hspace*{0.4em}\url{https://zenodo.org/records/20056072}%
}


\section{Dataset Construction}

RealDESED was collected in the context of a university machine learning course. The dataset comprises realistic domestic sound scenes with temporally precise annotations for 15 common domestic sound classes: \textit{bell ringing}, \textit{coffee machine}, \textit{cutlery and dishes}, \textit{door opening or closing}, \textit{footsteps}, \textit{keyboard typing}, \textit{keychain}, \textit{light switch}, \textit{microwave}, \textit{phone ringing or notification}, \textit{running water}, \textit{toilet flushing}, \textit{vacuum cleaner}, \textit{wardrobe or drawer opening or closing}, and \textit{window opening or closing}. Dataset creation consisted of three stages: data collection, annotation, and review, followed by a benchmark split into training, validation, and test sets.

\subsection{Data Collection}

Each participant was asked to collect eight to ten realistic domestic sound scenes between 15 and 35 seconds long. Recordings were made on consumer devices (smartphones, tablets, portable microphones, etc.) to reflect realistic home conditions.
Submitted files were converted to a unified waveform format during preprocessing.

The collection protocol encouraged balanced target-class coverage, multi-event and polyphonic scenes, and realistic acoustic conditions while avoiding unrealistically complex recordings. Natural background and non-target sounds, such as traffic noise, appliance hum, and chair movement were allowed and encouraged, provided the target events remained clearly audible. To increase acoustic diversity, collectors recorded scenes using \textit{static} and \textit{mobile} device placement, with the device placed on a surface or carried by the participant, respectively.
Recordings containing speech were not permitted.

Each recording was accompanied by structured metadata, including clip-level labels for audible target and non-target sounds, recording device, device placement, recording environment, and a short free-text scene description summarizing the sequence of actions and sound events. Non-target sounds, recording devices, and recording environments were not restricted to fixed vocabularies, whereas device placement was specified as either \textit{static} or \textit{mobile}. The metadata was later used for analysis and to support the annotation process. Collectors selected either CC0 or CC-BY as the license for their recordings.

\subsection{Data Annotation}

The recordings were annotated by 645 annotators using Label Studio Enterprise~\cite{labelstudio}. Each annotator was assigned at least 20 clips from a pool of 5,710 audio files. The assignment included the annotator's own recordings when available, supplemented with randomly selected clips from other collectors.

Annotators used an interface displaying the audio waveform, time-aligned spectrogram, and user-provided metadata. The metadata, including scene descriptions and target-class tags, was provided only as contextual guidance because it could be incomplete or inaccurate.

For each recording, annotators identified all audible events belonging to the 15 target classes and assigned each event a class label and temporal region. Overlapping events were allowed, while repeated events of the same class were annotated as separate regions if separated by a perceptually distinct pause. Non-target events were ignored.

In total, the annotation procedure produced 64,430 submitted annotations, with multiple annotations available for most recordings.

\subsection{Annotation Review}
\label{subsec:review}

The review process served as a qualitative quality-control step for the annotations. Each reviewer was assigned five to ten annotated recordings, and each reviewed recording had at least two independent annotations. This allowed reviewers to compare annotator decisions, identify inconsistencies, and correct the annotations where necessary.

For each recording, reviewers inspected the audio, metadata, waveform, spectrogram, and submitted annotations to verify class labels, temporal boundaries, and annotation completeness. When necessary, they corrected class labels and temporal regions. Exact boundary agreement across all annotations was not required, as small onset and offset differences may reflect perceptual ambiguity. In a subsequent quality-control step, approximately 300 low-quality reviews were identified based on remaining disagreement between annotations and reassigned for review, further improving consistency.


\subsection{Data Split}

The collected dataset contains 5,710 audio files, partitioned into training, validation, and test sets by collector ID, ensuring all recordings from the same domestic environment were assigned to the same split. The training, validation, and test sets contain 3,704, 999, and 1,007 recordings, respectively. All validation and test recordings were reviewed according to the procedure outlined in Section~\ref{subsec:review}, while approximately one third of the training set contains reviewed annotations.

\section{Dataset Characteristics}
\label{sec:dataset_characteristics}

The following section summarizes RealDESED, including the characteristics of its recordings, metadata, and temporal annotations, as well as the effect of the annotation review.


\subsection{Collected Audio Recordings and Metadata}
The 5,710 recordings total 37.85 hours. Scene durations range from 12.03 to 40.64 seconds, averaging 23.86 seconds. Despite a few deviations from the 15--35-second target, most recordings follow the intended short domestic-scene format.

\subsubsection{Target sounds}
The creator-reported clip-level labels indicate that recordings contain between 1 and 5 target sounds, with an average of 2.80.
The most frequent labels are footsteps (2,686 files) and door open close (1,791). The least frequent labels are bell ringing (430) and coffee machine (312). 
Common metadata-level co-occurrences include door open close with footsteps (1,278 files) and cutlery dishes with running water (629). Less plausible combinations, such as coffee machine with toilet flushing (1), are rare.

\subsubsection{Non-target sounds and scene descriptions}
The metadata shows substantial non-target sound diversity: 2,849 recordings (49.89\%) contain at least one non-target label, spanning 1,734 normalized labels.
Files contain 0.72 non-target labels on average, ranging from 0 to 8.
The most frequent normalized non-target sounds are clothing rustle (215 occurrences) and chair movement (142). Free-text scene descriptions have an average length of 24.56 words, suggesting that most collectors provided informative descriptions.

\subsubsection{Recording environments}
We normalize 161 recording-environment strings into 65 categories. Kitchens are most common (1,686 recordings), followed by bedrooms (1,091) and hallways (1,071), while balconies, garages, and staircases appear only once. 
\subsubsection{Recording devices}
The 523 raw recording-device strings were mapped to 342 device tags of varying specificity (``mobile phone'' to ``Fairphone 5''). Most recordings were made with mobile phones, accounting for 5,242 files (91.80\%). Tablets, laptops, and other devices were less common, accounting for 173 (3.03\%), 172 (3.01\%), and 123 recordings (2.15\%), respectively. The metadata further indicates 3,086 recordings with static and 2,624 recordings with mobile device placement. 


\subsection{Annotations}
Before review, the collected dataset contains 64,430 non-aggregated events across all 5,710 files with a total length of 77.80 hours. On average, each file has 2.45 annotators who marked an average of 4.60 event regions per file.
When taking the union of annotated regions within each file, 77.97\% (29.51 hours) of the total readable audio duration are covered by at least one event annotation. 

\subsubsection{Class distribution}
The distribution of the annotated events is imbalanced. The most frequent class is footsteps, with 11,597 raw annotation events, followed by door open close (7,978). The least frequent classes are coffee machine (1,140) and vacuum cleaner (1,278). This pattern is consistent with the metadata target class tags.

\subsubsection{Agreement between target class tags and strong annotations}
Metadata target classes are well covered by temporal annotations: 15,726 of 16,008 file--target class pairs appear in at least one raw annotation for the same file (98.24\%). 
Annotators also marked classes that were not listed in the metadata target tags. 2,641 of 64,430 annotated events (4.99\%) are outside the metadata target tags, affecting 1,134 files (19.86\%). The most common added classes are footsteps (533) and door open close (282). 

\subsubsection{Annotator agreement}
For files with multiple annotators, class-set agreement is relatively high but not perfect. When ignoring temporal annotations and comparing only the set of classes marked as present in each file (similar to clip-level tags) the mean pairwise Jaccard similarity is 0.875. 
For cases where multiple annotators marked the same class in the same file, pairwise temporal agreement is also reasonably high. The mean intersection-over-union over full recordings is 0.694. The mean overlap divided by the shorter total annotated duration is 0.929, showing that annotators often mark overlapping regions even when their exact onset and offset boundaries differ.

\subsubsection{Overlap analysis}
At least one temporal overlap of different target classes occurs in 7,155 of 14,009 file--annotator pairs (51.07\%). Across individual annotators, 19.25\% of annotated time is covered by two or more simultaneous events, with up to 4 simultaneous events.
The most common overlapping class pairs reflect plausible domestic situations such as cutlery dishes with running water (974 file--annotator pairs) and keyboard typing with phone ringing (866). 

\subsection{Review}
The validation and test splits were completely reviewed, covering all 999 validation files and all 1,007 test files. In the training split, 1,108 of 3,704 files were reviewed, corresponding to 29.91\% of the training data. 
The review process substantially improves agreement between annotations: For reviewed annotations, the mean class-set Jaccard similarity increases from 0.875 to 0.994, meaning that after review most annotations agree exactly on the set of classes present in a file. The pairwise mean temporal intersection-over-union increases from 0.694 for raw unreviewed same-file same-class annotations to 0.862 for accepted reviewed annotations.  Similarly, the mean overlap over the shorter annotation increases from 0.929 to 0.975.

\section{Benchmark Experiments}
\label{sec:results}

We establish the first benchmark results on RealDESED using ATST-F~\cite{atstf}, a frame-level Audio Spectrogram Transformer pre-trained on AudioSet Strong~\cite{PretrainedSED} and fine-tuned on RealDESED. Beyond establishing a strong baseline, we investigate three aspects specific to RealDESED: annotation aggregation for multi-annotator labels, long-form inference on recordings exceeding the model's input length, and the influence of recording metadata on SED performance.

\begin{table}[t]
\centering
\caption{Label aggregation strategies. Mean $\pm$ std over three runs.}
\label{tab:label_aggregation}
\scriptsize
\begin{tabular}{c@{\hspace{2pt}}lccc}
\toprule
\textbf{\#} & \textbf{Method} & \textbf{PSDS1-M}$\uparrow$ & \textbf{PSDS1}$\uparrow$ & \textbf{PSDS2-M}$\uparrow$ \\
\midrule
1  & Random (Fixed)            & $.675\pm.003$ & $.503\pm.010$ & $.916\pm.002$ \\
2  & Random (Epoch)            & $.692\pm.004$ & $.521\pm.002$ & $.933\pm.001$ \\
\midrule
3  & Majority                  & $.688\pm.004$ & $.518\pm.004$ & $.924\pm.002$ \\
4  & Intersection              & $.681\pm.007$ & $.511\pm.012$ & $.904\pm.001$ \\
5  & Union                     & $.673\pm.004$ & $.497\pm.002$ & $.925\pm.002$ \\
6  & Collector                 & $.682\pm.002$ & $.507\pm.002$ & $.920\pm.003$ \\
\midrule
7  & Uniform Soft              & $.690\pm.004$ & $.519\pm.005$ & $\mathbf{.933\pm.002}$ \\
8  & Weighted Soft ($\alpha$=16) & $.696\pm.007$ & $.532\pm.007$ & $.930\pm.003$ \\
\midrule
9  & Majority + Reviewed       & $.696\pm.006$ & $\mathbf{.533\pm.009}$ & $.927\pm.001$ \\
10 & Weighted + Reviewed       & $\mathbf{.697\pm.005}$ & $\mathbf{.533\pm.004}$ & $.929\pm.004$ \\
\bottomrule
\end{tabular}
\end{table}

\begin{table*}[t]
\centering
\caption{Comparison of post-processing methods. Results are averaged over three independent runs.}
\label{tab:postprocessing}
\scriptsize
\setlength{\tabcolsep}{2.5pt}
\begin{tabular}{lccc@{\hspace{4mm}}ccccccccccccccc}
\toprule
& \multicolumn{3}{c}{Overall} & \multicolumn{15}{c}{Class-wise PSDS1} \\
\cmidrule(lr){2-4}\cmidrule(l){5-19}
Method
& PSDS1-M
& PSDS1
& PSDS2-M
& Bell
& Coffee
& Cutlery
& Door
& Steps
& Typing
& Keychain
& Light
& Micro
& Phone
& Water
& Toilet
& Vacuum
& Ward.
& Window \\
\midrule

Raw
& 0.662
& 0.487
& 0.875
& 0.749
& 0.809
& 0.544
& 0.406
& 0.461
& 0.872
& 0.627
& 0.398
& 0.716
& 0.721
& 0.875
& 0.781
& 0.897
& 0.455
& 0.623
\\

Median Filter
& 0.693
& 0.528
& 0.903
& 0.748
& 0.827
& 0.594
& 0.465
& 0.565
& 0.866
& 0.659
& 0.404
& \textbf{0.787}
& 0.751
& 0.903
& 0.773
& 0.949
& 0.471
& 0.631
\\

cSEBBs
& \textbf{0.731}
& \textbf{0.573}
& \textbf{0.925}
& \textbf{0.789}
& \textbf{0.861}
& \textbf{0.637}
& \textbf{0.512}
& \textbf{0.584}
& \textbf{0.902}
& \textbf{0.706}
& \textbf{0.442}
& 0.784
& \textbf{0.768}
& \textbf{0.925}
& \textbf{0.860}
& \textbf{0.968}
& \textbf{0.530}
& \textbf{0.692}
\\

\bottomrule
\end{tabular}
\end{table*}

\subsection{Experimental Setup}

Training is performed on randomly sampled 10\,s crops from the training recordings, with frame-level labels generated at 25\,Hz. Since recordings exceed the model input length, different crops are sampled across epochs. Input preprocessing follows the original ATST-F setup~\cite{atstf}. The model is optimized using AdamW~\cite{adamw} with a peak learning rate of $4\times10^{-5}$, cosine learning-rate scheduling, Mixup~\cite{mixup}, frequency warping~\cite{atstf}, and binary cross-entropy loss. 

For long-form inference, recordings are processed using overlapping 10\,s windows with a default hop size of 5\,s. Unless stated otherwise, predictions are merged using triangular weighting (floor 0.3) and post-processed using a median filter with a window size of 360\,ms. These settings are varied in the corresponding experiments below.

Performance is evaluated using PSDS1 and PSDS2 scores~\cite{psds1_ebbers}. We report standard PSDS1 score ($\alpha_\mathrm{CT}=0$, $\alpha_\mathrm{ST}=1$) and macro-averaged PSDS1 and PSDS2 scores.

\subsection{Annotation Aggregation}
\label{subsec:annotation_aggregation}

RealDESED provides multiple annotations per recording, while approximately one third of the training set additionally contains reviewed annotations. We compare different annotation aggregation strategies during training (Table~\ref{tab:label_aggregation}).

\textit{Random (Fixed)} (Row \#1) establishes the single-annotator baseline by selecting one annotation per recording before training. In contrast, \textit{Random (Epoch)} (\#2) samples one annotation per recording in every epoch, substantially improving performance across all metrics. Simple aggregation methods, including majority vote (\#3), intersection (\#4), and union (\#5), consistently underperform this strategy. Row \#6 uses only annotations from the recording collector. Although this outperforms the fixed single-annotator baseline (\#1), it remains inferior to the best multi-annotator strategies.

We next replace hard label aggregation with soft frame-wise targets (\#7 and \#8). Let $y_i \in \{0,1\}^{C\times T}$ denote the binary frame-wise labels of annotator $i$, $C$ is the number of sound classes, $T$ the number of frames, and $N$ the number of annotators. For clarity, we omit the recording index throughout the following equations. Uniform soft labels are obtained by averaging the frame-wise annotations over the $N$ annotators, preserving annotator disagreement instead of collapsing it into hard labels. As shown in Row \#7, this approach outperforms hard aggregation strategies and achieves the highest PSDS2 Macro score across all strategies.

A natural extension is to weight annotators according to their annotation quality. We compute the frame-wise macro Dice agreement between pairs of annotators ($i$ and $j$),

\begin{equation}
D(y_i,y_j)
=
\frac{1}{|\mathcal{C}'|}
\sum_{c\in\mathcal{C}'}
\frac{2\langle y_i^{(c)},y_j^{(c)}\rangle}
{\|y_i^{(c)}\|_1+\|y_j^{(c)}\|_1},
\end{equation}

where $\mathcal{C}'$ denotes the set of active classes, $\langle \cdot,\cdot \rangle$ the inner product, and $\|\cdot\|_1$ the $\ell_1$-norm. For each annotator, their resulting Dice scores are averaged over all other annotators and recordings to obtain an annotator quality score $s_i$. Final soft labels are then computed as

\begin{equation}
w_i
=
\frac{s_i^\alpha}
{\sum_{j=1}^{N}s_j^\alpha},
\qquad
\tilde{y}
=
\sum_{i=1}^{N}w_i y_i,
\end{equation}

where $\alpha=16$ is selected on the validation set. As shown in Row~\#8, quality-weighted soft labels consistently outperform uniform averaging and achieve the best PSDS1 scores among methods without incorporating manually reviewed annotations.

Finally, Rows~\#9 and \#10 evaluate the benefit of incorporating the reviewed training annotations available for one third of the training split. The reviewed annotations replace the corresponding raw annotations, while the remaining recordings continue to use majority voting or weighted soft labels (based on \#3 and \#8), respectively. Reviewed annotations yield a larger PSDS1 gain for majority voting than for weighted soft labels. The comparatively small improvement for the latter suggests that annotator-aware weighting already produces higher-quality targets, making this automatic aggregation strategy competitive with human review. Overall, effectively leveraging multi-annotator annotations improves PSDS1 Macro from .675 to .697.

\subsection{Post-Processing}
\label{subsec:postprocessing}

Frame-wise SED predictions require post-processing to obtain temporally consistent event predictions. Building on the best label aggregation strategy (Row~\#10 in Table~\ref{tab:label_aggregation}) in terms of PSDS1 score, which uses a 360\,ms median filter, we optimize the post-processing strategy. We compare raw predictions, per-class median filtering, and cSEBBs~\cite{sebb}, with hyperparameters optimized on the validation set.

Table~\ref{tab:postprocessing} summarizes the results. Both post-processing methods substantially outperform the raw predictions, highlighting the importance of temporal modeling. Median filtering improves the PSDS1 Macro score from .662 to .693, but class-wise optimization provides no additional benefit and tends to overfit the validation set. cSEBBs achieves the best overall performance with .731 and is therefore used in all subsequent experiments.

Class-wise performance varies considerably across sound events. Sustained events such as \textit{vacuum cleaner} or \textit{running water} are detected reliably, whereas short-duration events including \textit{light switch} or \textit{door open close} remain the most challenging. cSEBBs improves performance for nearly all classes, demonstrating the benefit of strong temporal modeling.

\subsection{Long-Form Modeling}

\label{subsec:long-form}
Since RealDESED recordings exceed the 10\,s input length of the baseline model, inference is performed using overlapping sliding windows. Table~\ref{tab:longform} compares different window aggregation methods, triangular weighting functions, and hop sizes.

\begin{table}[t]
\centering
\caption{Influence of long-form inference parameters.}
\label{tab:longform}
\scriptsize
\renewcommand{\arraystretch}{0.95}
\begin{tabular}{llccc}
\toprule
\textbf{Category} & \textbf{Setting} & \textbf{PSDS1-M}$\uparrow$ & \textbf{PSDS1}$\uparrow$ & \textbf{PSDS2-M}$\uparrow$ \\
\midrule

\multirow{2}{*}{Aggregation}
& Average & \textbf{.731} & \textbf{.573} & \textbf{.925} \\
& Maximum & .721 & .565 & .910 \\

\midrule

\multirow{4}{*}{Filter floor}
& 0.0 & .703 & .551 & .881 \\
& 0.3 & \textbf{.731} & \textbf{.573} & \textbf{.925} \\
& 0.6 & .730 & \textbf{.573} & .917 \\
& 1.0 & .726 & .569 & .915 \\

\midrule

\multirow{4}{*}{Hop size (s)}
& 2.5 & \textbf{.736} & \textbf{.579} & .919 \\
& 5.0 & .731 & .573 & \textbf{.925} \\
& 7.5 & .718 & .561 & .910 \\
& 10.0 & .711 & .553 & .904 \\

\bottomrule
\end{tabular}
\end{table}

Average aggregation consistently outperforms maximum aggregation. Likewise, triangular weighting improves over uniform averaging (floor = 1.0), with a floor of 0.3 achieving the best overall performance. Finally, smaller hop sizes improve detection performance by increasing overlap between neighboring windows, although the gain from reducing the hop size from 5\,s to 2.5\,s is modest relative to the additional computational cost. We therefore use average aggregation, a triangular filter floor of 0.3, and a hop size of 5\,s.

\subsection{Metadata Analysis}
\label{subsec:metadata_analysis}

\begin{figure}[t]
    \centering
    \includegraphics[width=\linewidth]{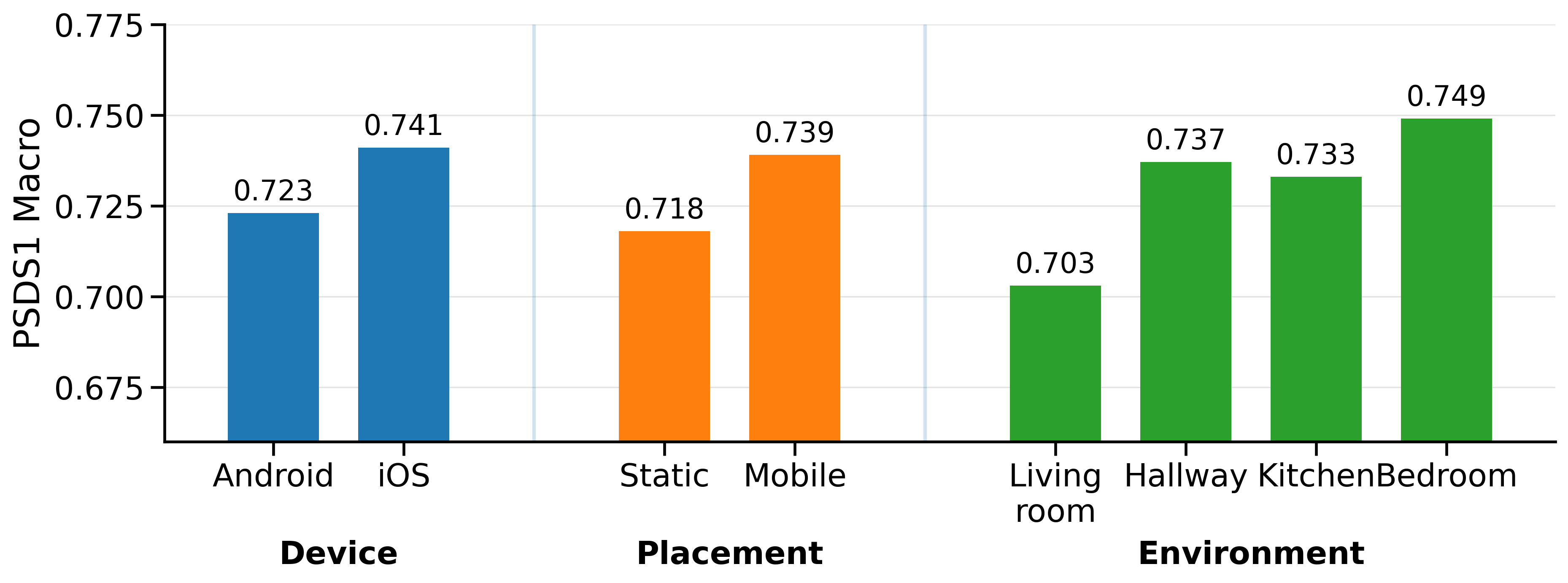}
    \caption{Macro-averaged PSDS1 over three runs grouped by recording device, device placement, and recording environment.}
    \label{fig:metadata}
\end{figure}

A key advantage of RealDESED is the availability of rich recording metadata, enabling analyses of model performance across different recording conditions. To illustrate the potential of the provided metadata, Figure~\ref{fig:metadata} reports macro-averaged PSDS1 grouped by recording device, device placement, and recording environment. For the device analysis, we compare Android smartphones against Apple devices (iPhones and iPads), excluding heterogeneous device categories such as laptops and other recording hardware.

The baseline exhibits meaningful performance differences across all three metadata dimensions. iOS devices outperform Android devices, mobile recordings outperform static recordings, and performance also varies considerably across recording environments. These results demonstrate that recording conditions can have a substantial impact on SED performance, motivating more detailed investigations into factors such as device characteristics, recording setup, and environment-specific event distributions. Such analyses highlight the potential of RealDESED as a benchmark for metadata-aware learning, domain adaptation, and device-robust SED.

\section{Conclusion}

We presented RealDESED, a new benchmark for domestic sound event detection comprising 5,710 real-world recordings collected by 652 participants in their homes. Compared to existing SED datasets, RealDESED provides recordings captured directly in target environments, temporally precise multi-annotator annotations, reviewed evaluation labels, and rich recording metadata. We established a strong transformer-based baseline and demonstrated the importance of annotation quality, temporal post-processing, and long-form inference, while highlighting the influence of recording conditions on SED performance. By publicly releasing the dataset, baseline implementation, and complete data collection and annotation protocol, we hope to provide a valuable resource for the community and a realistic benchmark for developing robust sound event detection systems that bridge the gap between current SED benchmarks and real-world deployment.




\clearpage
\bibliographystyle{IEEEtran}
\bibliography{refs}

@string{icassp = "Proc. ICASSP"}

@string{interspeech = "Proc. Interspeech"}

@string{waspaa = "Proc. WASPAA"}

@string{dcase = "Proc. DCASE"}

@string{ieee-acm-taslp = "IEEE/ACM Trans. Audio, Speech, Lang. Process."}

@string{iclr = "Proc. ICLR"}

@string{ieee-spm = "IEEE Signal Process. Mag."}

@string{ieee-tbme = "IEEE Trans. Biomed. Eng."}

@inproceedings{radhakrishnan2005audio,
  title        = {Audio analysis for surveillance applications},
  author       = {Radhakrishnan, Regunathan and Divakaran, Ajay and Smaragdis, A},
  booktitle    = waspaa,
  pages        = {158--161},
  year         = {2005},
  organization = {IEEE}
}

@article{debes2016monitoring,
  author       = {Christian Debes and
                  Andreas Merentitis and
                  Sergey Sukhanov and
                  Maria E. Niessen and
                  Nikolaos Frangiadakis and
                  Alexander Bauer},
  title        = {Monitoring Activities of Daily Living in Smart Homes: Understanding
                  human behavior},
  journal      = ieee-spm,  
  volume       = {33},
  number       = {2},
  pages        = {81--94},
  year         = {2016},
  _url         = {https://doi.org/10.1109/MSP.2015.2503881},
  _doi         = {10.1109/MSP.2015.2503881},
  _timestamp   = {Fri, 26 May 2017 22:53:35 +0200},
  _biburl      = {https://dblp.org/rec/journals/spm/DebesMSNFB16.bib},
  _bibsource   = {dblp computer science bibliography, https://dblp.org}
}

@article{zigel2009method,
  author       = {Yaniv Zigel and
                  Dima Litvak and
                  Israel Gannot},
  title        = {A Method for Automatic Fall Detection of Elderly People Using Floor
                  Vibrations and Sound - Proof of Concept on Human Mimicking Doll Falls},
  journal      = ieee-tbme,
  volume       = {56},
  number       = {12},
  pages        = {2858--2867},
  year         = {2009},
  _url         = {https://doi.org/10.1109/TBME.2009.2030171},
  _doi         = {10.1109/TBME.2009.2030171},
  _timestamp   = {Wed, 02 Sep 2020 13:24:36 +0200},
  _biburl      = {https://dblp.org/rec/journals/tbe/ZigelLG09.bib},
  _bibsource   = {dblp computer science bibliography, https://dblp.org}
}

@article{homesound,
  author       = {Rosa Maria Alsina{-}Pag{\`{e}}s and
                  Joan Navarro and
                  Francesc Al{\'{\i}}as and
                  Marcos Herv{\'{a}}s},
  title        = {homeSound: Real-Time Audio Event Detection Based on High Performance
                  Computing for Behaviour and Surveillance Remote Monitoring},
  journal      = {Sensors},
  volume       = {17},
  number       = {4},
  pages        = {854},
  year         = {2017},
  _url         = {https://doi.org/10.3390/s17040854},
  _doi         = {10.3390/S17040854},
  _timestamp   = {Wed, 14 Nov 2018 10:46:17 +0100},
  _biburl      = {https://dblp.org/rec/journals/sensors/Alsina-PagesNAH17.bib},
  _bibsource   = {dblp computer science bibliography, https://dblp.org}
}

@inproceedings{audioset,
  author       = {Jort F. Gemmeke and
                  Daniel P. W. Ellis and
                  Dylan Freedman and
                  Aren Jansen and
                  Wade Lawrence and
                  R. Channing Moore and
                  Manoj Plakal and
                  Marvin Ritter},
  title        = {Audio Set: An ontology and human-labeled dataset for audio events},
  booktitle    = icassp,
  pages        = {776--780},
  _publisher   = {{IEEE}},
  year         = {2017},
  _url         = {https://doi.org/10.1109/ICASSP.2017.7952261},
  _doi         = {10.1109/ICASSP.2017.7952261},
  _timestamp   = {Wed, 16 Oct 2019 14:14:52 +0200},
  _biburl      = {https://dblp.org/rec/conf/icassp/GemmekeEFJLMPR17.bib},
  _bibsource   = {dblp computer science bibliography, https://dblp.org}
}

@inproceedings{audioset_strong,
  author       = {Shawn Hershey and
                  Daniel P. W. Ellis and
                  Eduardo Fonseca and
                  Aren Jansen and
                  Caroline Liu and
                  R. Channing Moore and
                  Manoj Plakal},
  title        = {The Benefit of Temporally-Strong Labels in Audio Event Classification},
  booktitle    = icassp,
  pages        = {366--370},
  _publisher   = {{IEEE}},
  year         = {2021},
  _url         = {https://doi.org/10.1109/ICASSP39728.2021.9414579},
  _doi         = {10.1109/ICASSP39728.2021.9414579},
  _timestamp   = {Thu, 08 Jul 2021 17:13:36 +0200},
  _biburl      = {https://dblp.org/rec/conf/icassp/HersheyEFJLMP21.bib},
  _bibsource   = {dblp computer science bibliography, https://dblp.org}
}

@inproceedings{desed_turpault,
  author       = {Nicolas Turpault and
                  Romain Serizel and
                  Justin Salamon and
                  Ankit Parag Shah},
  _editor      = {Michael I. Mandel and
                  Justin Salamon and
                  Daniel P. W. Ellis},
  title        = {Sound Event Detection in Domestic Environments with Weakly Labeled
                  Data and Soundscape Synthesis},
  booktitle    = dcase,
  pages        = {253--257},
  year         = {2019},
  _url         = {http://dcase.community/documents/workshop2019/proceedings/DCASE2019Workshop\_Turpault\_44.pdf},
  _timestamp   = {Mon, 03 Jan 2022 17:42:19 +0100},
  _biburl      = {https://dblp.org/rec/conf/dcase/TurpaultSSS19.bib},
  _bibsource   = {dblp computer science bibliography, https://dblp.org}
}

@inproceedings{desed_serizel,
  author       = {Romain Serizel and
                  Nicolas Turpault and
                  Ankit Parag Shah and
                  Justin Salamon},
  title        = {Sound Event Detection in Synthetic Domestic Environments},
  booktitle    = icassp,
  pages        = {86--90},
  _publisher   = {{IEEE}},
  year         = {2020},
  _url         = {https://doi.org/10.1109/ICASSP40776.2020.9054478},
  _doi         = {10.1109/ICASSP40776.2020.9054478},
  _timestamp   = {Sun, 25 Oct 2020 23:13:32 +0100},
  _biburl      = {https://dblp.org/rec/conf/icassp/SerizelTSS20.bib},
  _bibsource   = {dblp computer science bibliography, https://dblp.org}
}

@inproceedings{scaper,
  author       = {Justin Salamon and
                  Duncan MacConnell and
                  Mark Cartwright and
                  Peter Li and
                  Juan Pablo Bello},
  title        = {Scaper: {A} library for soundscape synthesis and augmentation},
  booktitle    = waspaa,
  pages        = {344--348},
  _publisher   = {{IEEE}},
  year         = {2017},
  _url         = {https://doi.org/10.1109/WASPAA.2017.8170052},
  _doi         = {10.1109/WASPAA.2017.8170052},
  _timestamp   = {Sun, 04 Aug 2024 19:38:16 +0200},
  _biburl      = {https://dblp.org/rec/conf/waspaa/SalamonMCLB17.bib},
  _bibsource   = {dblp computer science bibliography, https://dblp.org}
}

@misc{labelstudio,
  title={{Label Studio}: Data labeling software},
  _url={https://github.com/HumanSignal/label-studio},
  note={Open source software available from https://github.com/HumanSignal/label-studio},
  author={
    Maxim Tkachenko and
    Mikhail Malyuk and
    Andrey Holmanyuk and
    Nikolai Liubimov},
  year={2020-2025},
}

@article{maestro_real,
  author       = {Irene Mart{\'{\i}}n{-}Morat{\'{o}} and
                  Annamaria Mesaros},
  title        = {Strong Labeling of Sound Events Using Crowdsourced Weak Labels and
                  Annotator Competence Estimation},
  journal      = ieee-acm-taslp,
  volume       = {31},
  pages        = {902--914},
  year         = {2023},
  _url         = {https://doi.org/10.1109/TASLP.2022.3233468},
  _doi         = {10.1109/TASLP.2022.3233468},
  _timestamp   = {Thu, 27 Jul 2023 08:18:02 +0200},
  _biburl      = {https://dblp.org/rec/journals/taslp/MartinMoratoM23.bib},
  _bibsource   = {dblp computer science bibliography, https://dblp.org}
}

@inproceedings{KongMil,
  author       = {Qiuqiang Kong and
                  Yong Xu and
                  Wenwu Wang and
                  Mark D. Plumbley},
  title        = {Audio Set Classification with Attention Model: {A} Probabilistic Perspective},
  booktitle    = icassp,
  pages        = {316--320},
  _publisher   = {{IEEE}},
  year         = {2018},
  _url         = {https://doi.org/10.1109/ICASSP.2018.8461392},
  _doi         = {10.1109/ICASSP.2018.8461392},
  _timestamp   = {Thu, 05 Jan 2023 07:59:59 +0100},
  _biburl      = {https://dblp.org/rec/conf/icassp/Kong0WP18.bib},
  _bibsource   = {dblp computer science bibliography, https://dblp.org}
}

@inproceedings{XuAttentionPooling,
  author       = {Yong Xu and
                  Qiuqiang Kong and
                  Wenwu Wang and
                  Mark D. Plumbley},
  title        = {Large-Scale Weakly Supervised Audio Classification Using Gated Convolutional
                  Neural Network},
  booktitle    = icassp,
  pages        = {121--125},
  _publisher   = {{IEEE}},
  year         = {2018},
  _url         = {https://doi.org/10.1109/ICASSP.2018.8461975},
  _doi         = {10.1109/ICASSP.2018.8461975},
  _timestamp   = {Thu, 05 Jan 2023 07:59:59 +0100},
  _biburl      = {https://dblp.org/rec/conf/icassp/0004KWP18.bib},
  _bibsource   = {dblp computer science bibliography, https://dblp.org}
}

@inproceedings{SchmidDCASE24Task4,
  author       = {Schmid, Florian and
                  Primus, Paul and
                  Morocutti, Tobias and
                  Greif, Jonathan and
                  Widmer, Gerhard},
  title        = {Multi-Iteration Multi-Stage Fine-Tuning of Transformers for Sound Event Detection with Heterogeneous Datasets},
  booktitle    = dcase,
  pages        = {141--145},
  year         = {2024},
  _address     = {Tokyo, Japan},
  _month       = {October},
  _abstract    = {A central problem in building effective sound event detection systems is the lack of high-quality, strongly annotated sound event datasets. For this reason, Task 4 of the DCASE 2024 challenge proposes learning from two heterogeneous datasets, including audio clips labeled with varying annotation granularity and with different sets of possible events. We propose a multi-iteration, multi-stage procedure for fine-tuning Audio Spectrogram Transformers on the joint DESED and MAESTRO Real datasets. The first stage closely matches the baseline system setup and trains a CRNN model while keeping the pre-trained transformer model frozen. In the second stage, both CRNN and transformer are fine-tuned using heavily weighted self-supervised losses. After the second stage, we compute strong pseudo-labels for all audio clips in the training set using an ensemble of fine-tuned transformers. Then, in a second iteration, we repeat the two-stage training process and include a distillation loss based on the pseudo-labels, achieving a new single-model, state-of-the-art performance on the public evaluation set of DESED with a PSDS1 of 0.692. A single model and an ensemble, both based on our proposed training procedure, ranked first in Task 4 of the DCASE Challenge 2024.}
}

@inproceedings{NamDCASE24Task4,
  author       = {Nam, Hyeonuk and
                  Min, Deokki and
                  Choi, Seungdeok and
                  Choi, Inhan and
                  Park, Yong-Hwa},
  title        = {Self Training and Ensembling Frequency Dependent Networks with Coarse Prediction Pooling and Sound Event Bounding Boxes},
  booktitle    = dcase,
  pages        = {96--100},
  year         = {2024},
  _address     = {Tokyo, Japan},
  _month       = {October},
  _abstract    = {To tackle sound event detection (SED), we propose frequency dependent networks (FreDNets), which heavily leverage frequency-dependent methods. We apply frequency warping and FilterAugment, which are frequency-dependent data augmentation methods. The model architecture consists of 3 branches: audio teacher-student transformer (ATST) branch, BEATs branch and CNN branch including either partial dilated frequency dynamic convolution (PDFD conv) or squeeze-and-Excitation (SE) with time-frame frequency-wise SE (tfwSE). To train MAESTRO labels with coarse temporal resolution, we applied max pooling on prediction for the MAESTRO dataset. Using best ensemble model, we applied self training to obtain pseudo label from DESED weak set, unlabeled set and AudioSet. AudioSet pseudo labels, filtered to focus on high-confidence labels, are used to train on DESED dataset only. We used change-detection-based sound event bounding boxes (cSEBBs) as post processing for ensemble models on self training and submission models. The resulting FreDNet was ranked 2nd in DCASE 2024 Challenge Task 4.}
}

@inproceedings{CornellDCASE24Task4,
  author       = {Cornell, Samuele and
                  Ebbers, Janek and
                  Douwes, Constance and
                  Martín-Morató, Irene and
                  Harju, Manu and
                  Mesaros, Annamaria and
                  Serizel, Romain},
  title        = {DCASE 2024 Task 4: Sound Event Detection with Heterogeneous Data and Missing Labels},
  booktitle    = dcase,
  pages        = {31--35},
  year         = {2024},
  _address     = {Tokyo, Japan},
  _month       = {October},
  _abstract    = {The Detection and Classification of Acoustic Scenes and Events Challenge Task 4 aims to advance sound event detection (SED) systems by leveraging training data with different supervision uncertainty. Participants are challenged in exploring how to best use training data from different domains and with varying annotation granularity (strong/weak temporal resolution, soft/hard labels), to obtain a robust SED system that can generalize across different scenarios. Crucially, annotation across available training datasets can be inconsistent and hence sound events of one dataset may be present but not annotated in an other one. As such, systems have to cope with potentially missing target labels during training. Moreover, as an additional novelty, systems are also evaluated on labels with different granularity in order to assess their robustness for different applications. To lower the entry barrier for participants, we developed an updated baseline system with several caveats to address these aforementioned problems. Results with our baseline system indicate that this research direction is promising and it is possible to obtain a stronger SED system by using diverse domain training data with missing labels compared to training a SED system for each domain separately.}
}

@inproceedings{PretrainedSED,
  author       = {Florian Schmid and
                  Tobias Morocutti and
                  Francesco Foscarin and
                  Jan Schl{\"{u}}ter and
                  Paul Primus and
                  Gerhard Widmer},
  title        = {Effective Pre-Training of Audio Transformers for Sound Event Detection},
  booktitle    = icassp,
  pages        = {1--5},
  _publisher   = {{IEEE}},
  year         = {2025},
  _url         = {https://doi.org/10.1109/ICASSP49660.2025.10888942},
  _doi         = {10.1109/ICASSP49660.2025.10888942},
  _timestamp   = {Tue, 01 Jul 2025 17:42:46 +0200},
  _biburl      = {https://dblp.org/rec/conf/icassp/SchmidMFSPW25.bib},
  _bibsource   = {dblp computer science bibliography, https://dblp.org}
}

@inproceedings{matsed,
  author       = {Pengfei Cai and
                  Yan Song and
                  Kang Li and
                  Haoyu Song and
                  Ian McLoughlin},
  _editor      = {Itshak Lapidot and
                  Sharon Gannot},
  title        = {{MAT-SED:} {A} Masked Audio Transformer with Masked-Reconstruction
                  Based Pre-training for Sound Event Detection},
  booktitle    = interspeech,
  _publisher   = {{ISCA}},
  year         = {2024},
  _url         = {https://doi.org/10.21437/Interspeech.2024-714},
  _doi         = {10.21437/INTERSPEECH.2024-714},
  _timestamp   = {Tue, 20 May 2025 17:13:53 +0200},
  _biburl      = {https://dblp.org/rec/conf/interspeech/Cai0LS024.bib},
  _bibsource   = {dblp computer science bibliography, https://dblp.org}
}

@inproceedings{pmam,
  author       = {Pengfei Cai and
                  Yan Song and
                  Nan Jiang and
                  Qing Gu and
                  Ian McLoughlin},
  title        = {Prototype based Masked Audio Model for Self-Supervised Learning of
                  Sound Event Detection},
  booktitle    = icassp,
  pages        = {1--5},
  _publisher   = {{IEEE}},
  year         = {2025},
  _url         = {https://doi.org/10.1109/ICASSP49660.2025.10889422},
  _doi         = {10.1109/ICASSP49660.2025.10889422},
  _timestamp   = {Tue, 05 Aug 2025 22:39:40 +0200},
  _biburl      = {https://dblp.org/rec/conf/icassp/CaiS00025.bib},
  _bibsource   = {https://dblp.org}
}

@article{jigsaw,
  author       = {Hyeonuk Nam and
                  Yong{-}Hwa Park},
  title        = {JiTTER: Jigsaw Temporal Transformer for Event Reconstruction for Self-Supervised
                  Sound Event Detection},
  journal      = {CoRR},
  volume       = {abs/2502.20857},
  year         = {2025},
  eprinttype   = {arXiv},
  eprint       = {2502.20857},
  _url         = {https://doi.org/10.48550/arXiv.2502.20857},
  _doi         = {10.48550/ARXIV.2502.20857},
  _timestamp   = {Wed, 26 Mar 2025 19:16:28 +0100},
  _biburl      = {https://dblp.org/rec/journals/corr/abs-2502-20857.bib},
  _bibsource   = {dblp computer science bibliography, https://dblp.org}
}

@article{atstf,
  author       = {Xian Li and
                  Nian Shao and
                  Xiaofei Li},
  title        = {Self-Supervised Audio Teacher-Student Transformer for Both Clip-Level
                  and Frame-Level Tasks},
  journal      = ieee-acm-taslp,
  volume       = {32},
  pages        = {1336--1351},
  year         = {2024},
  _url         = {https://doi.org/10.1109/TASLP.2024.3352248},
  _doi         = {10.1109/TASLP.2024.3352248},
  _timestamp   = {Thu, 29 Feb 2024 20:53:54 +0100},
  _biburl      = {https://dblp.org/rec/journals/taslp/LiSL24.bib},
  _bibsource   = {dblp computer science bibliography, https://dblp.org}
}

@inproceedings{adamw,
  author       = {Ilya Loshchilov and
                  Frank Hutter},
  title        = {Decoupled Weight Decay Regularization},
  booktitle    = iclr,
  _publisher   = {OpenReview.net},
  year         = {2019},
  _url         = {https://openreview.net/forum?id=Bkg6RiCqY7},
  _timestamp   = {Thu, 25 Jul 2019 14:26:04 +0200},
  _biburl      = {https://dblp.org/rec/conf/iclr/LoshchilovH19.bib},
  _bibsource   = {dblp computer science bibliography, https://dblp.org}
}

@inproceedings{mixup,
  author       = {Hongyi Zhang and
                  Moustapha Ciss{\'{e}} and
                  Yann N. Dauphin and
                  David Lopez{-}Paz},
  title        = {mixup: Beyond Empirical Risk Minimization},
  booktitle    = iclr,
  _publisher   = {OpenReview.net},
  year         = {2018},
  _url         = {https://openreview.net/forum?id=r1Ddp1-Rb},
  _timestamp   = {Thu, 25 Jul 2019 14:25:50 +0200},
  _biburl      = {https://dblp.org/rec/conf/iclr/ZhangCDL18.bib},
  _bibsource   = {dblp computer science bibliography, https://dblp.org}
}

@inproceedings{psds1_ebbers,
  author       = {Janek Ebbers and
                  Reinhold Haeb{-}Umbach and
                  Romain Serizel},
  title        = {Threshold Independent Evaluation of Sound Event Detection Scores},
  booktitle    = icassp,
  pages        = {1021--1025},
  _publisher   = {{IEEE}},
  year         = {2022},
  _url         = {https://doi.org/10.1109/ICASSP43922.2022.9747556},
  _doi         = {10.1109/ICASSP43922.2022.9747556},
  _timestamp   = {Mon, 03 Mar 2025 21:07:15 +0100},
  _biburl      = {https://dblp.org/rec/conf/icassp/EbbersHS22.bib},
  _bibsource   = {dblp computer science bibliography, https://dblp.org}
}

@inproceedings{sebb,
  author       = {Janek Ebbers and
                  Fran{\c{c}}ois G. Germain and
                  Gordon Wichern and
                  Jonathan Le Roux},
  _editor      = {Itshak Lapidot and
                  Sharon Gannot},
  title        = {Sound Event Bounding Boxes},
  booktitle    = interspeech,
  _publisher   = {{ISCA}},
  year         = {2024},
  _url         = {https://doi.org/10.21437/Interspeech.2024-2075},
  _doi         = {10.21437/INTERSPEECH.2024-2075},
  _timestamp   = {Tue, 20 May 2025 17:13:53 +0200},
  _biburl      = {https://dblp.org/rec/conf/interspeech/EbbersGWR24.bib},
  _bibsource   = {dblp computer science bibliography, https://dblp.org}
}







\end{document}